\documentclass[twocolumn,aps,prl,showpacs]{revtex4}
\usepackage{dcolumn}   

\usepackage{latexsym,amssymb,float}
\usepackage{setspace}
\usepackage{graphicx}
\usepackage{epsfig}
\usepackage{amsmath}
\usepackage{bm}

\begin{document}

\title{Non-equilibrium Transport in dissipative one-dimensional Nanostructures}
\author{Hui~Dai$^{1}$}
\author{Dirk K.~Morr$^{1,2}$}

\affiliation{$^{1}$ Department of Physics and James Franck
Institute, University of Chicago, Chicago, IL 60637, USA \\
$^{2}$University of Illinois at Chicago, Chicago, IL 60607, USA}
\date{\today}

\begin{abstract}
We study the non-equilibrium transport properties of a
one-dimensional array of dissipative quantum dots. Using the Keldysh
formalism, we show that the dots' dissipative nature leads to a
spatial variation of the chemical potential, which in disordered
arrays, breaks the invariance of the current, $I$, under bias
reversal. Moreover, the array's nanoscopic size results in an
algebraic low-temperature dependence of $I$. Finally, we show that a
local Coulomb interaction splits the dots' electronic levels,
resulting in a Coulomb blockade, which is softened with increasing
dissipation and array size.

\end{abstract}

 \maketitle

Arrays of metallic \cite{Tran05,Zab06} or semiconducting
\cite{Yu04,Hou08} quantum dots (QDs) have attracted significant
interest due to the unprecedented experimental control in assembling
such arrays, and the resulting ability to custom-design their
electronic structure and transport properties \cite{Shi00,Bel07}.
The continued miniaturization of such arrays, important for many
applications in quantum computation \cite{Loss98} and
nanoelectronics \cite{Shi00}, raises the important question of how
the non-equilibrium transport properties in nanoscopic arrays differ
from those in mesoscopic ones and how they evolve across different
length scales. Nanoscale arrays also provide a unique opportunity to
study how the local interplay between dissipation
\cite{Mal03,Jdi08}, disorder, and Coulomb interaction affects their
global transport properties. Investigating the evolution of these
properties across different length scales might also hold the key to
identifying the (yet unresolved) origin of the low-temperature
conductivity observed in mesoscopic QD arrays \cite{Tran05,Yu04}
which is ascribed either to some variable-range hopping (VRH)
\cite{Zab06,Yu04} or cotunneling \cite{Tran05,Bel05} mechanism.

\begin{figure}[h]
\includegraphics[width=8.cm]{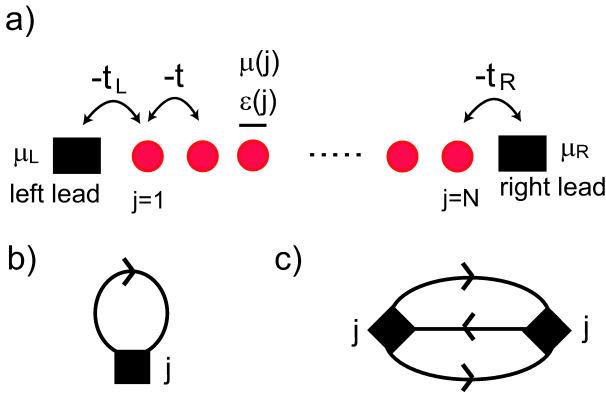}
\caption{(a) One-dimensional array of quantum dots coupled to two
leads. (b) First and (c) second order fermionic self-energy
corrections due to a local Coulomb interaction. \label{fig:Fig1}}
\end{figure}

In this Letter, we address the above questions by studying the
non-equilibrium transport properties of one-dimensional (1D),
nanoscale arrays of dissipative QDs, as schematically shown in
Fig.~\ref{fig:Fig1}(a).
Using the Keldysh Green's function formalism \cite{theory,Kel64}, we
demonstrate that the interplay between dissipation, Coulomb
interaction, and disorder gives rise to a series of novel
non-equilibrium quantum effects. In particular, the dots'
dissipative nature leads to a spatial variation of the chemical
potential, $\mu$, reflecting the array's resistance, which changes
qualitatively with increasing disorder, and breaks the invariance of
the current under bias reversal. Moreover, the array's nanoscopic
size yields an algebraic low-temperature dependence of the current
in disordered arrays, in contrast to the exponential scaling
observed in mesoscopic arrays \cite{Tran05,Yu04}. Finally, we
demonstrate that a local Coulomb interaction results in a splitting
of a dot's energy level, which continuously varies with the level's
occupation. This splitting gives rise to a Coulomb blockade in the
charge transport, which is softened with increasing dissipation
and/or size of the array. These results provide important new
insight into the complex non-equilibrium transport properties of
nanoscale systems.

We consider a one-dimensional array of $N$ quantum dots connected to
a left and right lead [Fig.~\ref{fig:Fig1}(a)] whose Hamiltonian is
given by ${\cal H}= {\cal H}_c +{\cal H}_{ph}+{\cal H}_{leads}$
where
\begin{eqnarray}
{\cal H}_c & = & \sum_{j,\sigma} \varepsilon(j) c^\dagger_{j,\sigma}
c_{j,\sigma} + U \sum_{j} \hat{n}_{j,\uparrow}\hat{n}_{j,\downarrow}
\nonumber \\  & & \hspace{-0.5cm} - \sum_{j,\sigma} t \;
c^\dagger_{j,\sigma} c_{j+1,\sigma} -\sum_{\sigma,\nu=R,L} t_\nu \;
c^\dagger_{n_\nu,\sigma} d_{\nu, \sigma}  + H.c.
\end{eqnarray}
Here $c^\dagger_{j,\sigma}$ and $d^\dagger_{\nu, \sigma}$ create an
electron with spin $\sigma$ on dot $j$ and lead $\nu$ ($\nu=R,L$),
respectively. We assume for simplicity that on each dot, there is
only a single electronic state with energy $\varepsilon(j)$ that is
relevant for charge transport, as is the case in semiconducting QDs
\cite{Jdi08}. $U$ represents the strength of the local (intra-dot)
Coulomb interaction, and $t$ is the hopping matrix element between
neighboring dots. $t_\nu$ describes the coupling of the array to the
leads with $n_{L (R)} = 1 (N)$. Finally, ${\cal H}_{ph}$ and ${\cal
H}_{leads}$ represent the local electron-phonon interaction on each
dot and the leads' electronic structure, respectively.

In order to study the array's non-equilibrium transport properties,
we consider a (symmetric) bias, $V=\mu_L-\mu_R$, across the array
arising from different chemical potentials in the left and right
leads with $\mu_R=-\mu_L$. The resulting current between dots $j$
and $(j+1)$ is given by \cite{Car71a}
\begin{equation}
I_{j,j+1}=-\frac{e}{\hbar} \; t
\intop_{-\infty}^{+\infty}\frac{d\omega}{2\pi}{\rm Re} \left[{\hat
F}_{j,j+1}(\omega)\right] \ .
\end{equation}
Here, ${\hat F}$ is the full Keldysh Green's functions matrix of the
entire array, which accounts for (electron-phonon and Coulomb)
interactions, electronic hopping between dots and the coupling to
the leads. Since scanning tunneling spectroscopy (STS) on QD arrays
\cite{Jdi08} reported that intra-dot interactions are larger than
the inter-dot electronic hopping \cite{Jdi08}, we include the former
first in the perturbative calculation of ${\hat F}$, in contrast to
the limit considered previously \cite{Mit04,Bih05}. Moreover, STS
experiments \cite{Mal03,Jdi08} also reported significant energy
broadening (i.e., dissipation) of the dots' electronic states
(likely arising from electron-phonon coupling) which is only weakly
temperature and energy dependent, and remains substantial even for
$T \rightarrow 0$ \cite{Jdi08}. We account for the dissipation by
introducing a (phenomenological) lifetime, $\Gamma^{-1}$, of the
dots' electronic states (we take $\varepsilon$ to include a possible
shift due to the electron-phonon interaction) \cite{Mit04,Bih05}.
The Green's function of a single (isolated) dot is then (the effects
of a Coulomb interaction are discussed below)
\begin{equation}
g_{r}^{-1}(\omega)=\omega-(\varepsilon-\mu)+i\Gamma  \ .
\label{eq:Gr}
\end{equation}
Including next the electronic hopping between the dots and the
coupling to the leads, we obtain
\begin{equation}
\hat{F} =
\left(1-\hat{g_{r}}\hat{t}\right)^{-1}\hat{f}\left(1-\hat{t}\hat{g_{a}}\right)^{-1}
\ . \label{eq:F}
\end{equation}
Here, $\hat{g}_{r,a}$ and $\hat{f}$ are the (diagonal) retarded,
advanced and Keldysh Greens function matrices describing the
decoupled ($t=t_\nu=0$) dots and leads, respectively, with
\begin{equation}
{\hat f}(\omega) = 2i \left[1-2 {\hat n_F}(\omega) \right] {\rm Im}
\left[ {\hat g}_r(\omega) \right] \ . \label{eq:f0}
\end{equation}
${\hat n_F}$ is a diagonal matrix containing the dots' and leads'
Fermi-distribution functions, and $\hat{t}$ is the symmetric hopping
matrix. While the results shown below were obtained in arrays with
$N=20$ dots, we found qualitatively similar behavior up to the
largest arrays ($N=150$) we studied. For concreteness, we take
$t=t_{R,L}=0.3E_{0}$ and $\Gamma=0.05E_{0}$, unless otherwise
specified.

\begin{figure}
\includegraphics[width=8.5cm]{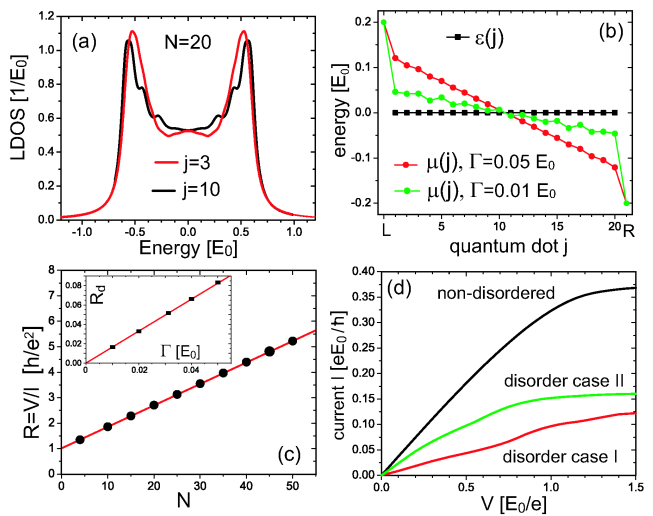}
\caption{QD array with $N=20$. (a) $N_j(\omega)$ in equilibrium with
$\varepsilon(j)=\mu(j)=0$. (b) $\mu(j)$, $\varepsilon(j)$ for
$V=0.4E_{0}/e$ and different $\Gamma$. $L$ and $R$ refer to the left
and right leads, respectively. (c) Resistance $R=V/I$ as a function
of $N$. Inset: $R_d$ as a function of $\Gamma$. (d) $IV$-curve for a
non-disordered array and two different disorder realizations with
$s=0.2 E_0$.} \label{fig:Fig2}
\end{figure}

In order to understand an array's transport properties, it is
instructive to first consider its (equilibrium) electronic
structure. To this end, we present in Fig.~\ref{fig:Fig2}(a) the
local density of states (LDOS), $N_j(\omega) = - {\rm Im} {\hat
G}^r_{jj}(\omega) /\pi$ where ${\hat G}_r
=\left(1-\hat{g_{r}}\hat{t}\right)^{-1} \hat{g_{r}}$ of a
non-disordered array with $N=20$ dots. The array's nanoscopic 1D
structure gives rise to a significant curvature of the band (see
discussion below), with the LDOS exhibiting the precursors of the
$1/\sqrt{E}$-divergence characteristic of an array with $N=\infty$.

We next discuss the array's non-equilibrium transport properties.
For a given bias $V$, charge conservation in the steady-state
requires that all currents $I_{j,j+1}$ be the same. For dissipative
QDs with $\Gamma \not = 0$, this requirement can only be satisfied
by allowing spatial variations in the local chemical potential
\cite{Bue86}, $\mu(j)$, of each dot [Fig.~\ref{fig:Fig2}(b)], which
enters in the calculation of $I_{j,j+1}$ through ${\hat f}(\omega)$
in Eqs.(\ref{eq:F}) and (\ref{eq:f0}). Hence, $\mu$ should be
interpreted as the electro-chemical potential and its difference
between two adjacent dots as the change in the electrostatic
potential due to the array's resistance.
As expected, the spatial variation of $\mu$ decreases with
decreasing $\Gamma$ and vanishes for a non-dissipative array with
$\Gamma \rightarrow 0$ \cite{Kel64}. A plot of the resistance,
$R=V/I$, as a function of $N$ [Fig.~\ref{fig:Fig2}(c)] reveals a
linear relationship $R=2R_I + R_d N $, with $R_I \approx
0.50\hbar/e^{2}$ and $R_d \approx 0.084\hbar/e^{2}$ being the
interface resistance and resistance per dot, respectively,
demonstrating that the array behaves as a series of resistors. Since
$R_I \gg R_d$, the change in $\mu$ at the interface is significantly
larger than that between dots, in agreement with
Fig.~\ref{fig:Fig2}(b). Moreover, $R_d \sim \Gamma$ (see inset)
demonstrates that it is the dissipative nature of the QDs that is
responsible for the array's resistance. Finally, a plot of the
$IV$-curve in Fig.~\ref{fig:Fig2}(d) shows ohmic behavior, $I \sim
V$, at low bias while $I$ becomes basically independent of $V$ once
$V$ exceeds the electronic band width, $W \approx 4t = 1.2 E_0$, of
the array.

\begin{figure}
\includegraphics[width=8.5cm]{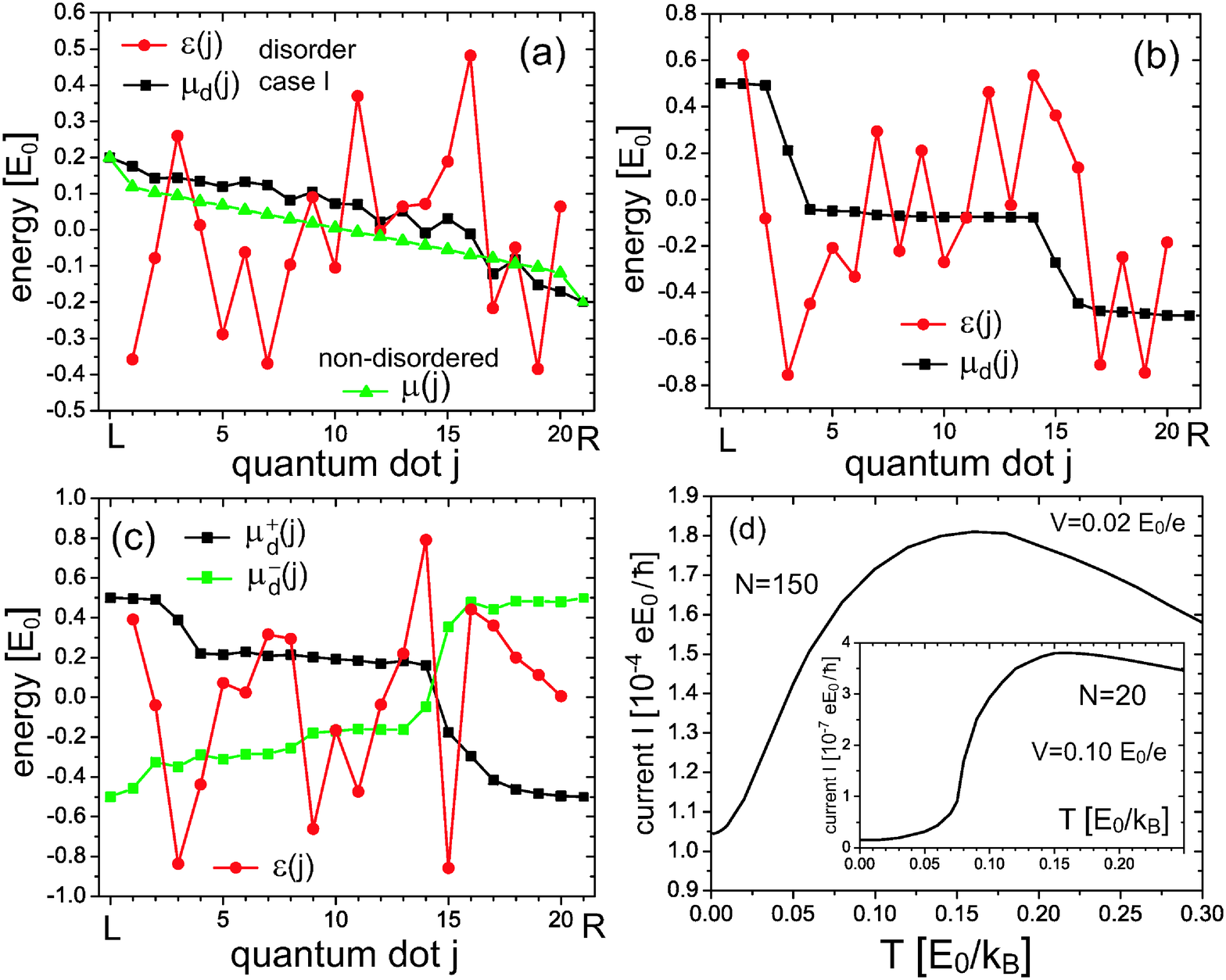}
\caption{$\varepsilon(j)$, $\mu(j)$ for (a) a non-disordered array
and disorder case $I$, and (b) for a strongly disordered array with
$t=\Gamma=0.01 E_0$, and $s=0.5 E_0$. (c) $\varepsilon(j)$,
$\mu^{\pm}_d(j)$ for a disordered array under bias reversal. (d)
$I(T)$ for an array with $N=150$ at intermediate disorder with
$s=t=0.3 E_0$ and $\Gamma = 0.05 E_0$. Inset: $I(T)$ for the
strongly disordered array of (b). } \label{fig:Fig4}
\end{figure}

Disorder in an array can emerge from non-uniformity in the dots'
spacing or size, leading to variations in $t$ and $\varepsilon(j)$,
respectively. Below, we consider the latter case, which has a more
profound effect on the array's transport properties, and take
$\varepsilon(j)$ to be Gaussian (disorder-) distributed with average
value $\langle \varepsilon(j) \rangle =0$ and standard deviation
$s$. The $IV$-curve for two realizations of intermediate disorder
($t \sim s$), case I and II with $s=0.2 E_0$, are plotted in
Fig.~\ref{fig:Fig2}(d) [$\varepsilon(j)$ for case I is shown in
Fig.~\ref{fig:Fig4}(a)].
While in both cases, the current is suppressed by disorder, the
extent of the suppression depends on the specific disorder
realization. The microscopic origin of this suppression lies in the
diminished hybridization of electronic states between neighboring
dots due to the differences in their energies, $\varepsilon(j)$. In
Fig.~\ref{fig:Fig4}(a) we present $\varepsilon(j)$ and $\mu_d(j)$
for disorder case I together with $\mu(j)$ for the non-disordered
array. Surprisingly, $\mu_d(j)$ does not differ significantly from
$\mu(j)$, and in particular, does not track the variations in
$\varepsilon(j)$. The reason for this robustness of $\mu_d$ is that
in the intermediate disorder case, electrons can tunnel even through
those dots where $|\varepsilon|$ is large. In contrast, in the
strong disorder limit, $t,\Gamma \ll s$, shown in
Fig.~\ref{fig:Fig4}(d), $\mu_d(j)$ exhibits step-like changes
between those neighboring dots where the variation in
$\varepsilon(j)$ is the largest. This is expected since large
spatial variations of $\varepsilon(j)$ indicate the regions of
largest resistance in the array, with concomitant large (potential)
changes in $\mu_d(j)$.

The interplay between disorder and dissipation gives rise to an
interesting novel quantum phenomenon in which the invariance of the
current's magnitude is broken under bias reversal. To demonstrate
this effect, we present in Fig.~\ref{fig:Fig4}(c) the chemical
potentials, $\mu^{\pm}_d(j)$, in a dissipative, disordered array for
two different bias $V_\pm=\pm E_0/e$. Due to the combination of
disorder and dissipation, $\mu^{+}_d$ and $\mu^{-}_d$ are not
related by a spatial symmetry, implying that the corresponding
currents are in general different: in Fig.~\ref{fig:Fig4}(c), the
current for positive bias $V_+$ is $I_+ = 0.0089  e E_0/\hbar$,
while for $V_-$ one finds $I_- = -0.0261 e E_0/\hbar$. We note that
the invariance is restored either in a non-disordered dissipative
array, or in a non-dissipative ($\Gamma = 0^+$) disordered array. In
the latter case, the current is independent of $\mu$ \cite{Kel64}
and $\mu^{\pm}_d(j) \equiv 0$ again satisfies the spatial symmetry.

The temperature dependence of $I$ is generally determined by both
disorder and interaction effects, with the latter being represented
by $\Gamma$. However, when disorder becomes stronger than
dissipation (i.e., $s \gtrsim \Gamma$), we find that the current is
predominantly determined by the former, and becomes essentially
independent of $\Gamma$. Thus, in calculating $I(T)$, we can neglect
the temperature dependence of interaction effects, i.e.,
$\Gamma(T)$, as was previously assumed in Refs.\cite{Mott68,Efr75}.
In the limit of intermediate disorder, $\Gamma \ll t \sim s$ , and
for $k_B T \ll s$, $I(T)$ possesses an algebraic temperature
dependence, $I(T)=I_0 + B T^2$, as shown in Fig.~\ref{fig:Fig4}(d)
for an array with $N=150$ dots. This algebraic dependence is the
lowest order (in $T$) contribution arising from the frequency
dependence of the Green's functions entering Eq.(\ref{eq:F}), which
in turn arises from the finite 1D structure of the array [this
frequency dependence is also present in the LDOS of
Fig.~\ref{fig:Fig2}(a)]. In the strong disorder limit, $\Gamma, t
\ll s$ (see inset of Fig.~\ref{fig:Fig4}(d)], the algebraic
temperature dependence of $I(T)$ at small $T$ is followed by a more
rapid increase at larger temperatures, which can be described by an
activated behavior over some limited temperature range. We note that
the absence of an exponential VRH temperature scaling
\cite{Mott68,Efr75}, is one of the characteristic hallmarks of
nanoscale systems. Not only would the algebraic contribution to
$I(T)$ dominate any exponential scaling for $T \rightarrow 0$, but
VRH scaling also assumes that a disordered system is characterized
by a single parameter, the localization length, and thus neglects
spatial fluctuations in the disorder. These fluctuations are,
however, highly relevant in nanoscale systems due to the lack of
disorder self-averaging, resulting in different $IV$-curves for
different disorder realizations with the same $s$
[Fig.~\ref{fig:Fig2}(d)]. Work is currently under way to identify
the characteristic size of the QD array at which disorder
self-averaging will lead to a crossover from algebraic to VRH
scaling of $I(T)$.

\begin{figure}
\includegraphics[width=8.5cm]{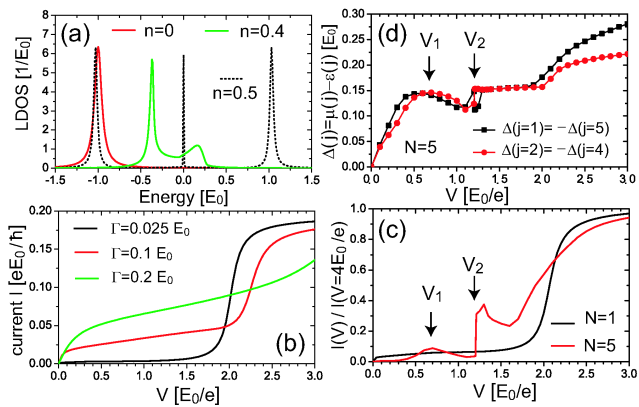}
\caption{(a) LDOS of a single dot with $\Gamma = 0.05 E_0$ for
$n_{j, \sigma}=0,0.4,0.5$. (b) $IV$-curve for an array with $N=1$
and different $\Gamma$. (c) Normalized $IV$-curve for arrays with
$N=1$ and $N=5$ dots. (d) $\Delta(j)=\mu(j)-\varepsilon(j)$ as a
function of $V$ for $N=5$.} \label{fig:Fig3}
\end{figure}

We next study the effects of a local Coulomb interaction by
including the first [Fig.~\ref{fig:Fig1}(b)] and second order
[Fig.~\ref{fig:Fig1}(c)] fermionic self-energy corrections in the
local Green's functions entering Eq.(\ref{eq:F}). Based on the
experimental STS results of Ref.~\cite{Jdi08}, we take the Coulomb
interaction to be weaker than the electron-phonon interaction, and
thus take the internal fermionic lines of the diagrams in
Figs.~\ref{fig:Fig1}(b) and (c) to be given by the dissipative
Green's function of Eq.(\ref{eq:Gr}). The first order diagram of
Fig.~\ref{fig:Fig1}(b) leads to an energy shift ${\bar
\varepsilon}_{\sigma}(j)=\varepsilon(j)+U n_{j,{\bar \sigma}}$ with
$n_{j,\sigma}$ being the electron occupation and $\sigma, {\bar
\sigma}$ being opposite spin quantum numbers. In contrast, the
second order diagram, for which an analytical form can be derived,
leads to an energy splitting which depends on $n_{j,\sigma}$: for a
completely empty or filled state, the splitting vanishes, while it
is the largest for a half-filled state. To demonstrate the evolution
of the dot's electronic structure with $n_{j,\sigma}$, we present in
Fig.~\ref{fig:Fig3}(a) the LDOS of a single dot [including the
self-energy corrections of Figs.~\ref{fig:Fig1}(b),(c)] for $n_{j,
\sigma}=0.0,0.4, 0.5$, and $U=2 E_0$ (the LDOS for occupation
$1-n_{j,\sigma}$ is obtained via $\omega \leftrightarrow -\omega$).
For $n_{j,\sigma}=0$ ($n_{j,\sigma}=1$), the dot possess a single,
doubly degenerate energy level at $\omega=-E_0$ ($\omega = + E_0$).
As $n_{j,\sigma}$ increases, this level shifts to larger energies,
and a splitting emerges. For $n_{j, \sigma}=0.5$, the state's
spectral weight is predominantly located at $\omega = \pm E_0$,
i.e., separated by $U=2 E_0$, while a small contribution of spectral
weight is found at $\omega=0$, which is a direct consequence of
$\Gamma \not = 0$, and hence vanishes for $\Gamma \rightarrow 0$.
This evolution of the LDOS with $n_{j,\sigma}$ is identical to that
of Ref.~\cite{Meir91} for $\Gamma = 0^+$, but qualitatively differs
from it for $\Gamma \not = 0$. The state's splitting, which is
analogous to the Coulomb interaction driven suppression of the LDOS
near the Fermi energy discussed by Efros and Shklovskii
\cite{Efr75}, is the origin of the Coulomb blockade observed in $IV$
curves. For a single QD with $\mu=0$ and $n_{j, \sigma}=0.5$
[Fig.~\ref{fig:Fig3}(b)], the dot's energy level is maximally split,
and $I$ is strongly suppressed until $V$ exceeds $U$, thus
manifesting the Coulomb blockade. The state's spectral weight at
$\omega=0$ (due to $\Gamma \not = 0$) leads to a non-zero $I$ even
for a $V<U$. The Coulomb blockade is softened both with increasing
$\Gamma$ [Fig.~\ref{fig:Fig3}(b)] as well as with increasing number
of dots $N$ in the array, as shown in Fig.~\ref{fig:Fig3}(c) for
$N=5$. In addition, in the latter case, new features appear for
$V<U$ which arise from changes in $\Delta(j)
=\mu(j)-\varepsilon(j)$, where $\Delta(j) =0$ corresponds to
$n_{j,\sigma}=0.5$ and thus a maximum splitting of the energy level.
As $\Delta(j)$ increases, this splitting is reduced, and $I$
increases (and vice versa) as follows from a comparison of
Figs.~\ref{fig:Fig3}(c) and (d). Therefore, the local maximum of $I$
at $V_1 \approx 0.7 E_0/e$ coincides with that of $\Delta(j)$, while
its sharp increase at $V_2 \approx 1.2 E_0$ reflects that of
$\Delta(j)$. For $V>V_2$, $I$ decreases slightly due to an
increasing separation between the dots' energy levels, but then
increases again as $V$ approaches the Coulomb gap, $U$.

In summary, we studied the non-equilibrium transport properties of a
1D array of dissipative QDs. We showed that the QDs' dissipative
nature leads to a spatial variation of $\mu(j)$, reflecting the
resistance of the array. In disordered arrays, this variation breaks
the invariance of the current under bias reversal while the array's
nanoscopic size results in an algebraic low-temperature dependence
of $I(T)$. Finally, we showed that a local Coulomb interaction gives
rise to a splitting of the dots' energy levels, and the emergence of
a Coulomb blockade, which is softened with increasing dissipation
and size of the array.

We would like to thank J. Figgins, I. Gruzberg, P. Guyot-Sionnest,
H. Jaeger, L. Kadanoff, A. Kamenev, and G.P. Parravicini for
stimulating discussions. D.K.M. would like to thank the Aspen Center
for Physics for its hospitality during the final stage of this
project. This work is supported by the U.S. Department of Energy
under Award No. DE-FG02-05ER46225 (D.K.M) and by NSF MRSEC under
Award No. 0820054 (H.D.).

\end{document}